\documentclass{IEEEtran}
\usepackage{cite}
\usepackage{amsmath,amssymb,amsfonts}
\usepackage{algorithmic}
\usepackage{graphicx}
\usepackage{textcomp}
\usepackage{algorithmic}
\usepackage{algorithm}
\usepackage{tabularx}
\usepackage{multirow}
\usepackage{diagbox}
\usepackage{multirow}
\usepackage{colortbl}
\usepackage{xcolor}
\usepackage{subfigure}
\pdfoutput=1
\usepackage{setspace}
\definecolor{headerblue}{RGB}{0, 91, 153}
\definecolor{lightgray}{RGB}{240, 240, 240}

\def\BibTeX{{\rm B\kern-.05em{\sc i\kern-.025em b}\kern-.08em
    T\kern-.1667em\lower.7ex\hbox{E}\kern-.125emX}}
\begin{document}
\title{Improving Beam Granularity Performance of Reconfigurable Refelctarray Radars via Spatial Quantization and Phase Quantization Approach}
\author{Xiaocun Zong, Fan Yang,  \IEEEmembership{Fellow, IEEE}, Shenheng Xu, \IEEEmembership{Member, IEEE}, and Maokun Li, \IEEEmembership{Fellow, IEEE}

\thanks{This work is supported by the National Key Research and Development Program of China under Grant No. 2023YFB3811501. (Corresponding author: Fan Yang.) }
\thanks{The authors are with the  Department of Electronic Engineering, State Key laboratory of Space Network and Communications, Tsinghua University, Beijing 100084, China (e-mail: fan\_yang@tsinghua.edu.cn). }}

\maketitle

\begin{abstract}
In this paper, the impacts of spatial quantization and phase quantization on the beam granularity characteristic of reconfigurable reflectarray (RRA) radars are systematically investigated. From the perspective of the difference beam, a theoretical analysis is conducted to derive the factors influencing beam granularity. To validate the theoretical findings, simulations are performed under various quantization scenarios: specifically, 1-bit, 2-bit, and 3-bit spatial quantization with 1-bit phase quantization, as well as 1-bit, 2-bit, and 3-bit phase quantization with 1-bit spatial quantization. The experimental results demonstrate that both spatial quantization and phase quantization effectively reduce beam granularity in reconfigurable reflectarray radars, thereby enhancing the angular resolution of the beam. These findings offer valuable insights and practical reference for beam-tracking applications in radar and communications.
\end{abstract}


\section{Introduction}
\label{sec:introduction}
\IEEEPARstart{B}{EAM} granularity, defined as the finest realizale increment between adjacent beam positions, is a critical parameter in the field of radar detection [1] and communication system [2], which determines the beam's precision tracking performance.

In reconfigurable reflectarrays, phase quantization results in a discrete phase distribution across the antenna array, leading to discrete beam steering positions and, consequently, discrete beam granularity. Minimizing beam coverage losses is critical for radar and communication systems to ensure robust detection performance, which can be achieved by reducing beam granularity. In discrete multi-target tracking, precise alignment of the beam direction with the predicted target position is essential. Coarse beam granularity degrades the signal-to-noise ratio (SNR), reduces tracking accuracy, and may lead to target loss. Thus, reducing beam granularity is crucial for enhancing system performance. However, few studies have focused on optimizing beam granularity in reconfigurable reflectarrays, and fundamental approaches to improving this characteristic during the design remain largely unexplored.

Drawing inspiration from phase quantization [3], the concept of spatial quantization is propsed. It defines how finely an array is divided. Higher spatial quantization accuracy results in finer subdivisions and smaller element size. This reduces proportion of a single element relative to the entire array. Similarly, improved phase quantization accuracy decreases the minimum phase increment per element. Consequently, the impact of a element’s phase change on the array diminishes. Thus it is hypothesized that spatial quantization and phase quantization can improve beam granularity performance. This forms the motivation and foundation of our research.
{
\setlength{\parskip}{0.3em}

\textbf{\textit{Define }}\(\mathit{S_N}\):$\left\{ \mathit{N\text{-}bit\ spatial\ quantization:}\ P_{\mathit{unit}} = \frac{\lambda}{2^N} \right\}$

\textbf{\textit{Define }}\(\mathit{P_N}\):$\left\{ \mathit{N\text{-}bit\ phase\ quantization:}\ \Delta\varphi_{\mathit{min}} = \frac{2\pi}{2^N} \right\}$

}
{
\setlength{\parskip}{0.3em}

This article is organized as follows: Section \uppercase\expandafter{\romannumeral2} explains the }basic composition and principle of reconfigurable reflectarray radar, as well as the related concepts of beam granularity and difference beam; Section \uppercase\expandafter{\romannumeral3} develops a theoretical framework, deriving the factors influencing beam granularity through an abstracted model; Section \uppercase\expandafter{\romannumeral4} validates the theoretical findings via simulation, demonstrating that spatial quantization and phase quantization enhance beam granularity performance; In section \uppercase\expandafter{\romannumeral5}, a law of how the beam granularity changes with the element position is further explored, and a theoretical formula with guiding significance is given.\uppercase\expandafter{\romannumeral6} provides a summary of the paper.

\section{Basic Principles of Reconfigurable Reflectarray Radar}
The basic structure of a reconfigurable reflectarray radar, as shown in Fig. 1, typically consists of two main components: a feed horn and a digitally controllable electromagnetic surface.
\begin{figure}[H]
	\centering
	\includegraphics[width=0.55\linewidth]{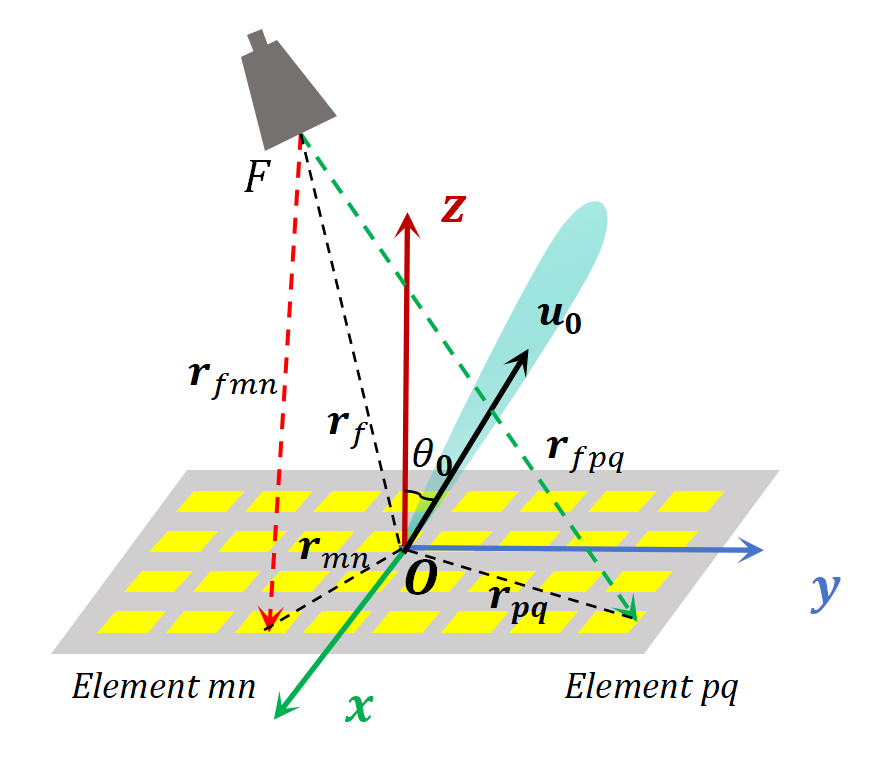}% 中括号中的为调节图片大小
	\caption{A general structure of reconfigurable reflectarray radar.}
\end{figure}

In order to achieve the focusing of the beam in the $u_0$ direction, ideally, the phase of each element on the array surface should form an equal phase surface in the direction perpendicular to $u_0$ after compensating the radiation of the feed source. Assuming that all elements on the surface are located in the far field of the feed source, and each element should provide a phase:

\begin{equation}
\varphi_{mn}^{req}=k\cdot (\vec {r}_{fmn}- u_0\cdot \vec {r}_{mn})+\varphi_{0}
\end{equation}

 For a reconfigurable reflectarray radar, the phase of the element is a discrete value. The continuous $\varphi_{mn}^{req}$ is quantified to the nearest quantized value $\varphi_{mn}$, allowing the desired beam to be synthesized effectively.

 In radar research, when analyzing scenes that are sensitive to angle changes, the radar's pattern is usually in the form of a difference beam. The change in zero depth position of difference beam is the granularity, and it is more obvious than the sum beam pattern. When designing the difference beam, the element phases of one half array are needed to flipped by 180°. The concept and phase distribution of difference beam can be seen in Fig. 2.
\begin{figure}[H]%调节图片位置，h：浮动；t：顶部；b:底部；p：当前位置
	\centering
	\includegraphics[width=0.85\linewidth]{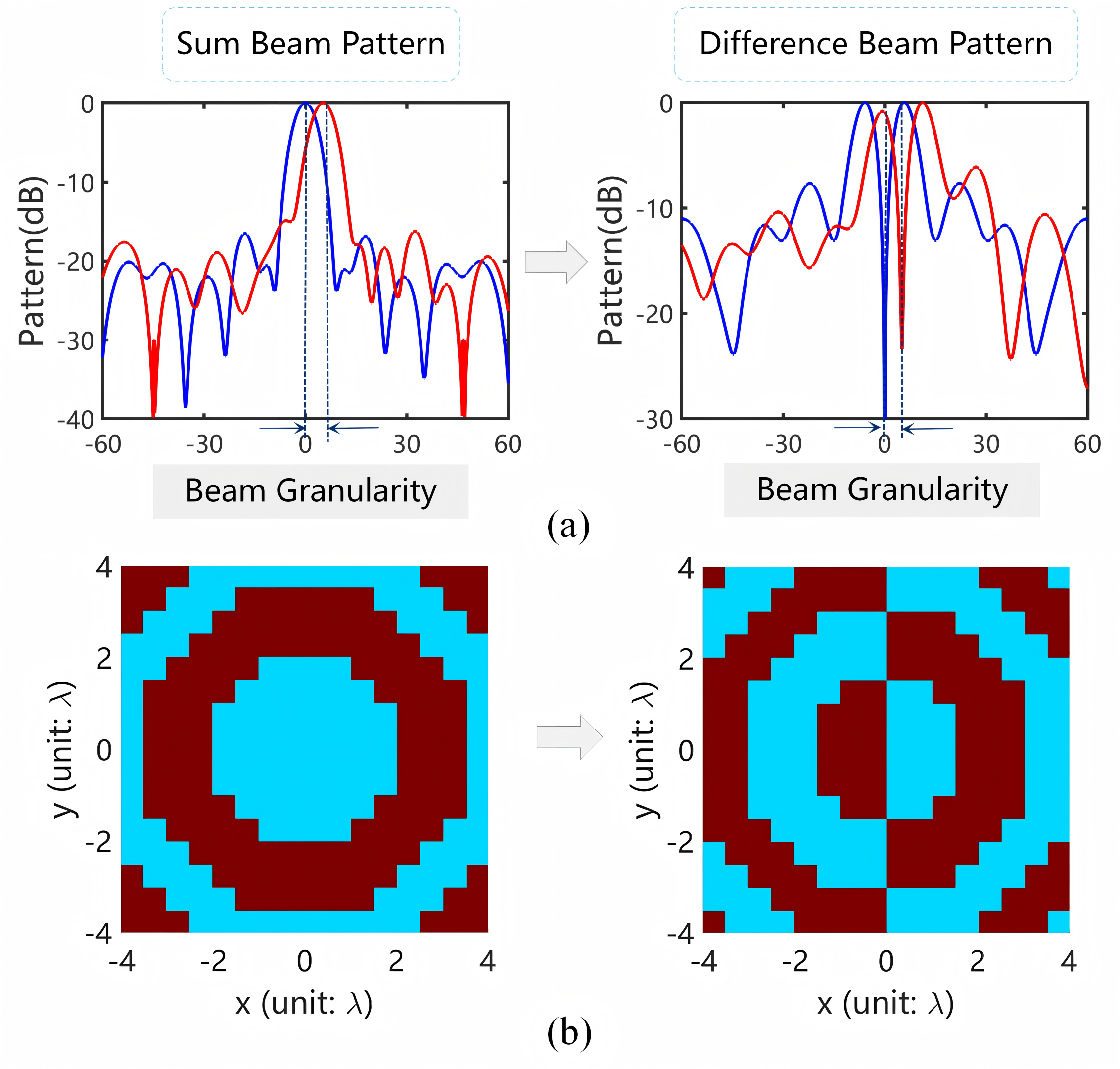}% 中括
	\caption{(a) The beam granularity concept of sum beam and difference beam; (b) The element phase distribution of sum beam and difference beam.}
\end{figure}

\section{Theoretical Derivation of Beam Granularity Calculation}

In order to facilitate the derivation of the beam granularity theoretical formula, it is assumed that all elements are omnidirectional and have the same amplitude [4]. The difference beam radiation pattern of the beam pointing to $(\theta_0,\varphi_0)$ can be expressed as:
\begin{equation}
\begin{aligned}
E_{\theta}(\theta,\varphi)=f_{element}\sum_{m=0}^{M/2-1}\sum_{n=0}^{N-1}&e^{j\frac{2\pi md}{\lambda}(sin\theta-sin\theta_0)}\\
&e^{j\frac{2\pi nd}{\lambda}(cos\theta sin\varphi-cos\theta_0sin\varphi_0)}\\
-f_{element}\sum_{m=M/2}^{M-1}\sum_{n=0}^{N-1}&e^{j\frac{2\pi md}{\lambda}(sin\theta-sin\theta_0)}\\
&e^{j\frac{2\pi nd}{\lambda}(cos\theta sin\varphi-cos\theta_0sin\varphi_0)}
\end{aligned}
\end{equation}
where $f_{element}$ is the element radiation pattern, assuming $f_{element}=1$ , let \begin{equation}
\left\{
\begin{aligned}
X &= \left( \frac{2\pi d}{\lambda} \right) \left( \sin\theta - \sin\theta_0 \right) \\
Y &= \left( \frac{2\pi d}{\lambda} \right) \left( \cos\theta \sin\varphi - \cos\theta_0 \sin\varphi_0 \right)
\end{aligned}
\right.
\end{equation} 

then equation (2) can be further simplified to:
\begin{equation}
E_{\theta}(\theta,\varphi)=\frac{ 2 \left( sin\frac{MX}{4}\right)^2} {sin\frac{X}{2}}\frac{sin\frac{NY}{2}}{sin\frac{Y}{2}}e^{j\left(\frac{M-1}{2}X-\frac{\pi}{2}\right)}e^{j\frac{N-1}{2}Y}
\end{equation}

Thus, the elevation difference beam amplitude pattern function is obtained:
\begin{equation}
|E_{\theta}(\theta,\varphi)|=\left| \frac{ 2 \left( sin\frac{MX}{4}\right)^2} {sin\frac{X}{2}}\frac{sin\frac{NY}{2}}{sin\frac{Y}{2}}\right|
\end{equation}

A slight phase shift applied to any element’s phase shifter state—either an increase or decrease by the minimum phase shift step—will cause a small change in the position of the zero-depth point of the difference beam pattern. This shift corresponds to the minimum beam granularity achievable under the given beam pointing condition.To determine the minimum beam granularity, the process is as follows:

\begin{flushleft}
\textit{
1). Calculate the new difference beam pattern function after the beam pointing changes; \\
2). Obtain the corresponding amplitude change value of the difference beam with respect to the angular change; \\
3). Divide the amplitude change value by the slope of the difference beam pattern at the zero-depth point. The resulting value represents the angular change in the pattern, which is the minimum beam granularity.
}
\end{flushleft}

Taking the elevation difference beam pattern as an example, when a phase shifter on the array produces a  phase change, the pattern function is:
\begin{equation}
\begin{aligned}
    \Delta E_{\theta}(\theta,\varphi)=&E_{\theta}(\theta,\varphi)-(-1)^le^{jmX}e^{jnY}\\
    &+(-1)^le^{jmX-\Delta \varphi}e^{jnY}
    \end{aligned}
\end{equation}

In the formula: $l$ is 0 when $m<\frac{M}{2}$, and is $1$ when $m>\frac{M}{2}$. When $\theta\rightarrow\theta_0, \varphi\rightarrow\varphi_0, X\rightarrow0, Y\rightarrow0$ ,
\begin{equation}
    |\Delta E_{\theta}(\theta,\varphi)|=\sqrt{2-2cos\Delta\varphi}
\end{equation}

Taking the partial derivative of the difference beam pattern in equation (5), the slope of the difference beam pattern can be calculated. Since $\Delta\theta$ is very small, it can be assumed that the slope of the difference pattern at $\theta_0+\Delta\theta$ and $\theta_0$ is approximately equal at zero depth. When $\theta\rightarrow\theta_0, \varphi\rightarrow\varphi_0, X\rightarrow0, Y\rightarrow0$. Using the same method, we can get::
\begin{equation}
    \frac{\partial|E_{\theta}(\theta,\varphi)|}{\partial\theta}=\frac{NM^2\pi d}{2\lambda}cos\theta_0
\end{equation}

According to equations (7) and (8), the minimum beam granularity formula of difference beam can be obtained:

\begin{equation}
\Delta\theta_{min} = \frac{2\sqrt{2(1-\cos\Delta\varphi)}\lambda}{MN^2\pi d\cos\theta_0} 
\end{equation}

In summary, through theoretical analysis of the difference beam pattern function, the minimum beam granularity calculation formula for the difference beam pattern is obtained.

\section{Simulation Verification of Spatial Quantization and Phase Quantization to Improve Beam Granularity}

\subsection{Verify the Correctness of the Theory}
To verify the correctness of the theory, the array beam granularity with equal-amplitude omnidirectional element under different quantization conditions is simulated. From Table I, it can be seen that under the condition of equal-amplitude omnidirectional elements, the simulation results are roughly consistent with the theoretical results. It is worth pointing out that at this time, the elements at different positions have different beam granularity due to different phase quantization errors; after optimization, we obtained a set of more reliable data.
\begin{table}[H]
    \centering
    \begingroup
    \footnotesize  % 设置字体变小
    \renewcommand{\arraystretch}{1.1} % 调整行高
    \caption{Comparison Between Theory and Simulation at 0° Beam}
    \label{tab:comparison}
    \begin{tabularx}{\linewidth}{>{\centering\arraybackslash}p{0.12\linewidth} *{5}{>{\centering\arraybackslash}X}}
        \hline
        \multirow{2}{*}{\textbf{}} 
        & \multicolumn{5}{c}{Beam Granularity (deg)} \\
        \cline{2-6}
        & \textbf{$P_1S_1$} & \textbf{$P_1S_2$} & \textbf{$P_1S_3$} & \textbf{$P_2S_1$} & \textbf{$P_3S_1$} \\
        \hline
        Theory & 0.0356 & 0.0089 & 0.0023 & 0.0252 & 0.0136 \\
        Simulation & 0.0315 & 0.0086 & 0.0029 & 0.0257 & 0.0115 \\
        \hline
    \end{tabularx}
    \endgroup
\end{table}

For reconfigurable reflectarray radar, the process of finding the minimum beam granularity is slightly different due to the different element amplitudes. For theoretical simulations with uniform amplitude distribution, it is only necessary to find the element with the minimum phase error, while for reflectarray radars with non-uniform amplitude distribution, it is necessary to consider the combined effects of amplitude and phase. The detailed process is shown in Fig. 3. The numerical differences and reasons will be discussed later.

\begin{figure}[h]%调节图片位置，h：浮动；t：顶部；b:底部；p：当前位置
	\centering
	\includegraphics[width=0.99\linewidth]{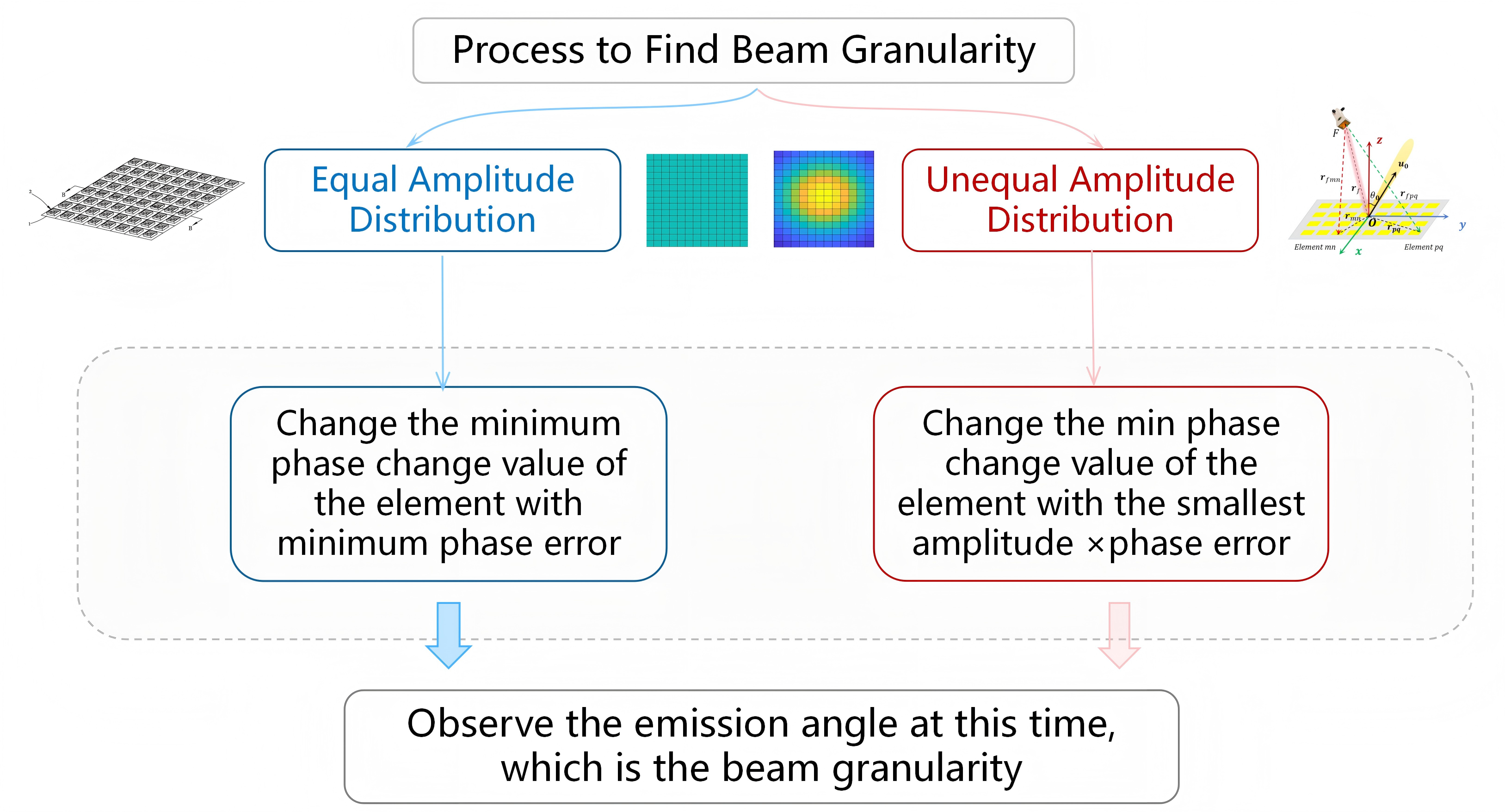}% 中括
	\caption{Process to find beam granularity of uniform amplitude distribution and non-uniform amplitude distribution.}
\end{figure}

\subsection{Reconfigurable Reflectarray Sptail Quantization Effect}
To investigate the effect of spatial quantization on beam granularity in reconfigurable reflectarray radars, the beam granularity of configurations $P_1S_1$, $P_1S_2$, and $P_1S_3$ is analyzed under a fixed aperture size. 

The edge element in Fig. 4 green box, which exhibits a reduced amplitude effect, produces the smallest beam granularity. Simulation results, summarized in Table II and Fig. 4, indicate beam granularities of 0.0229°, 0.0037°, and 0.0006° respectively. The theoretical model partially explains the observed variations in beam granularity. Notably, the beam granularity of the reconfigurable reflectarray radar is smaller than the theoretical prediction, as elements with lower amplitudes have a reduced impact on the overall array performance.

\begin{figure*}[ht]
    \centering
    \begin{subfigure}
        \centering
        \includegraphics[width=0.8\textwidth]{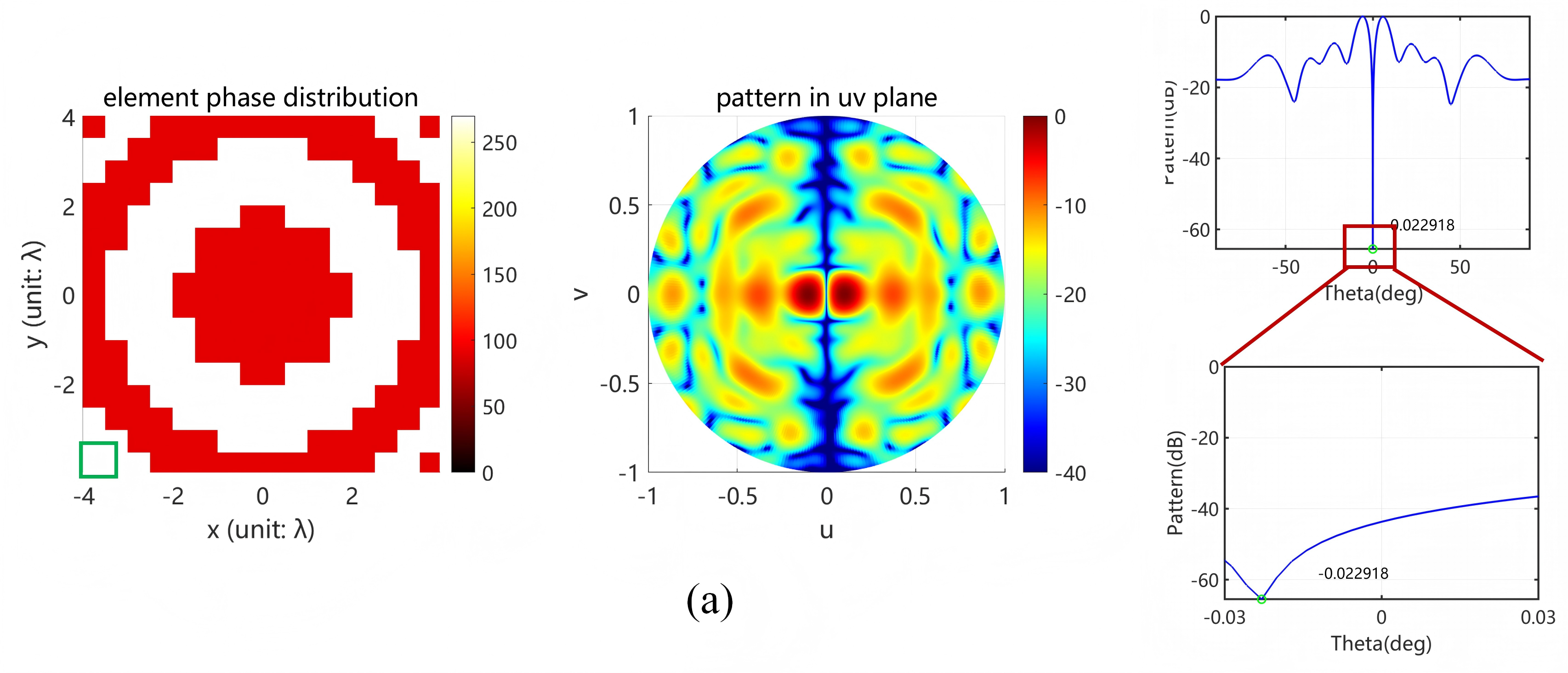}
        %\caption{First image}
        %\label{fig:sub1}
    \end{subfigure}
    
    \begin{subfigure}
        \centering
        \includegraphics[width=0.8\textwidth]{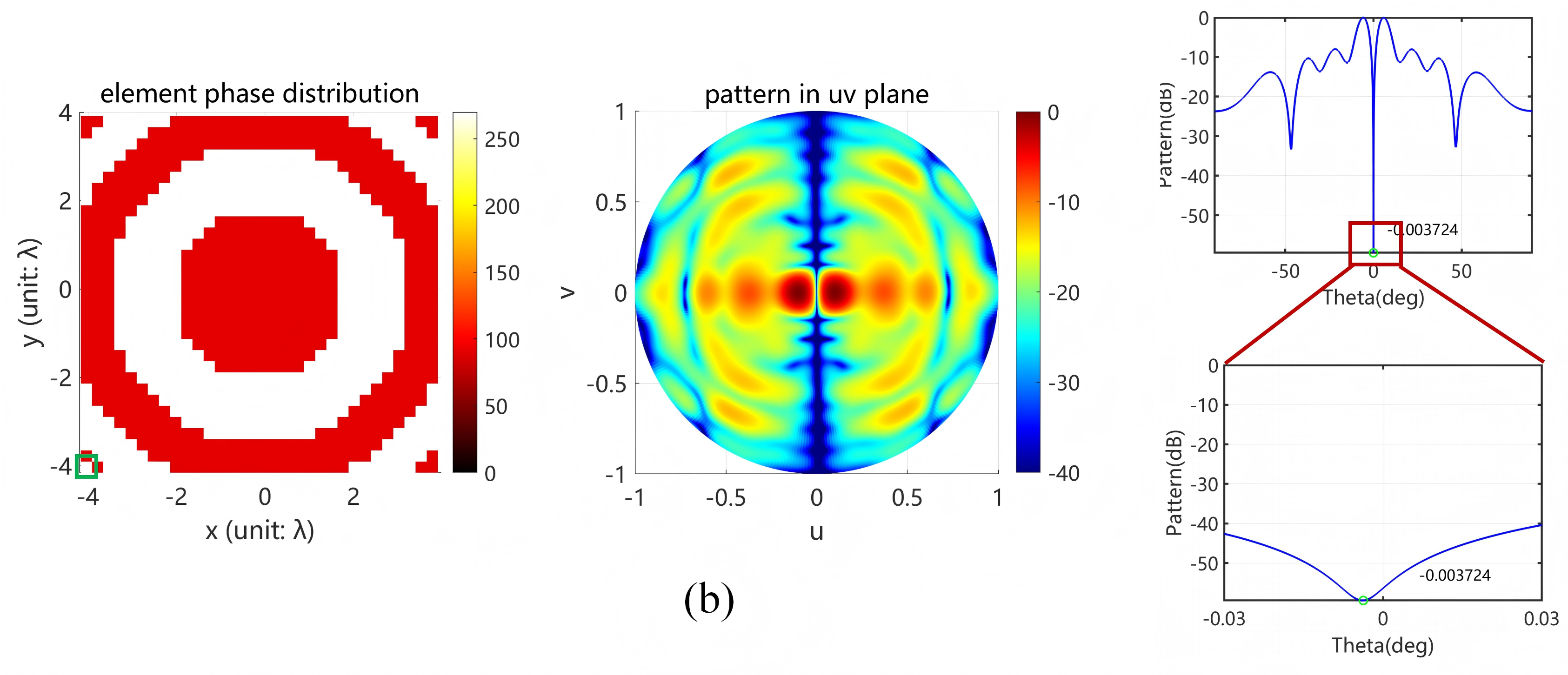}
        %\caption{Second image}
        %\label{fig:sub2}
    \end{subfigure}
 
    \begin{subfigure}
        \centering
        \includegraphics[width=0.8\textwidth]{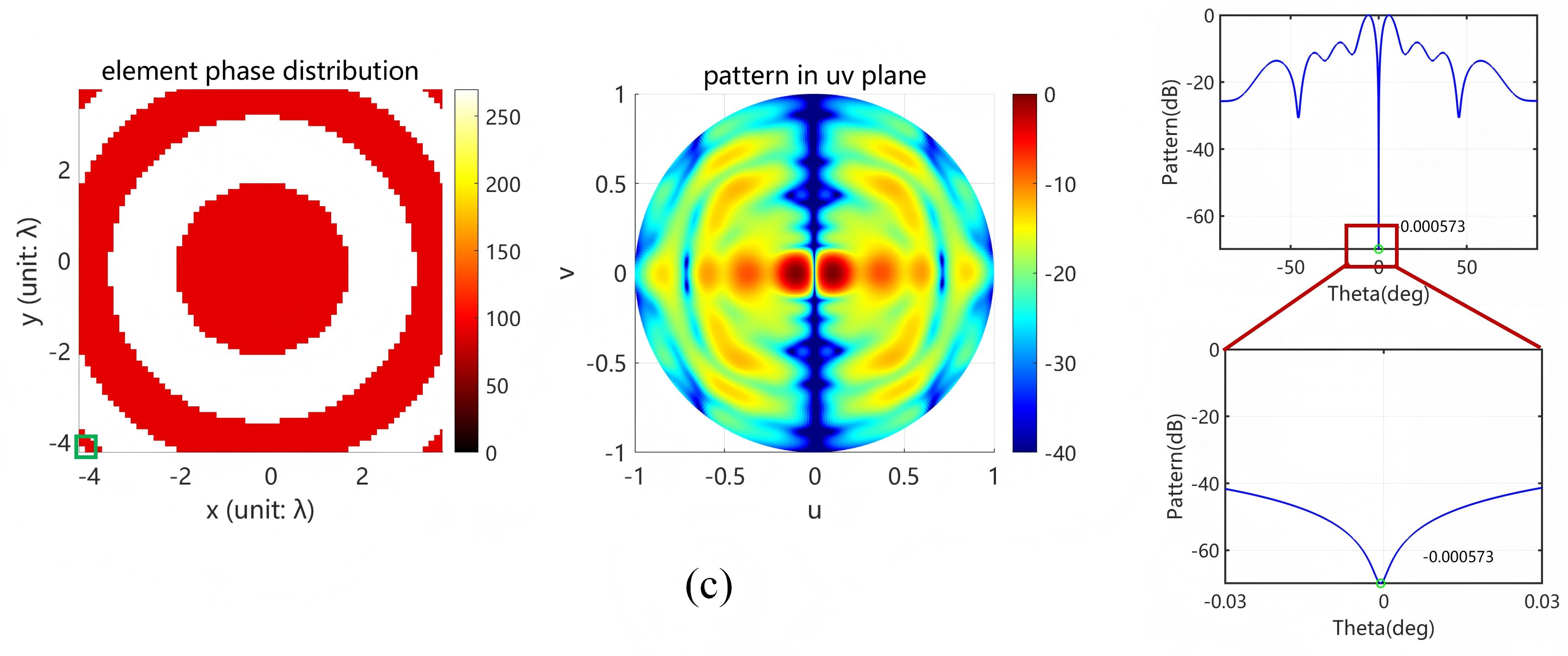}
        %\caption{Third image}
        %\label{fig:sub3}
    \end{subfigure}
    
    \caption{Beam granularity under different spatial quantization bits:(a)$P_1S_1$, (b)$P_1S_2$, (c)$P_1S_3$}
    \label{fig:overall}
\end{figure*}

\subsection{Reconfigurable Reflectarray Phase Quantization Effect}
According to the previous method, the beam granularity under different phase quantization conditions can be studied. The results are presented in Table III. As the number of phase quantization bits increases, the beam granularity incrementally decreases. This is because a smaller quantization phase will have a smaller effect.
\begin{table}[h]
    \centering
    \begingroup
    \footnotesize  % 设置字体变小
    \renewcommand{\arraystretch}{1.1} % 调整行高
    \caption{Beam Granularity under Different Spatial Quantization Conditions}
    \label{tab:comparison}
    \begin{tabular}{p{0.3\columnwidth} *{3}{p{0.15\columnwidth}}} % 第一列0.35\columnwidth，第2-4列均分
        \hline
        \multirow{2}{*}{\textbf{}} 
        & \multicolumn{3}{c}{\parbox{0.45\columnwidth}{\centering {Beam Granularity (deg)}}} \\
        \cline{2-4}
        & \textbf{\centering $P_1S_1$} & \textbf{\centering $P_1S_2$} & \textbf{\centering $P_1S_3$} \\
        \hline
        \parbox{0.3\columnwidth}{\centering Theoretical Analysis} & 0.0356 & 0.0089 & 0.0023 \\
        \parbox{0.3\columnwidth}{\centering Theory Simulation} & 0.0315 & 0.0086 & 0.0029 \\
        \parbox{0.3\columnwidth}{\centering RRA Simulation}  & 0.0229 & 0.0037 & 0.0006 \\
        \hline
    \end{tabular}
    \endgroup
\end{table}
\begin{table}[h]
    \centering
     \begingroup
    \footnotesize  % 设置字体变小
    \renewcommand{\arraystretch}{1.1} % 调整行高
    \caption{Beam Granularity under Different Phase Quantization Conditions}
    \label{tab:comparison}
    \begin{tabular}{p{0.3\columnwidth} *{3}{p{0.15\columnwidth}}} % 第一列0.35\columnwidth，第2-4列均分
        \hline
        \multirow{2}{*}{\textbf{}} 
        & \multicolumn{3}{c}{\parbox{0.45\columnwidth}{\centering {Beam Granularity (deg)}}} \\
        \cline{2-4}
        & \textbf{\centering $P_1S_1$} & \textbf{\centering $P_2S_1$} & \textbf{\centering $P_3S_1$} \\
        \hline
        \parbox{0.3\columnwidth}{\centering Theoretical Analysis} & 0.0356 & 0.0252 & 0.0136 \\
        \parbox{0.3\columnwidth}{\centering Theory Simulation} & 0.0315 & 0.0257 & 0.0115 \\
        \parbox{0.3\columnwidth}{\centering RRA Simulation} & 0.0229 & 0.0115 & 0.0057 \\
        \hline
    \end{tabular}
    \endgroup
\end{table}

In addition, full-wave simulation is also conducted to verify this conclusion. The results of full-wave simulation also reveal the change law, but due to the influence of factors such as the different element losses in different states and the error of the actual phase, there is a slight difference between the experimental data and the simulation. Anyway, it has been proved that spatial quantization and phase quantization can effectively improve beam granularity performance.

\section{Further Discussion}
To identify the element position that minimizes beam granularity in reconfigurable reflectarray radars, the influence of elements at various positions on beam granularity was thoroughly investigated. Specifically, a formula was derived to quantify the contribution of element amplitude and phase to beam granularity.
\begin{equation}
    Target\ Element = min\{A_{mn}cos(\varphi_{mn}+\Delta\varphi_{min}-\varphi_{mn}^{req})\}
\end{equation}
among them, $A_{mn}$ represents the contribution of the element amplitude; $\varphi_{mn}$ is the phase after being quantized, $\Delta\varphi_{min}$ is the minimum quantized phase, and $\varphi_{mn}^{req}$ is the theoretical phase in Eq. (1). These three phases represent the phase error after changing the minimum phase; therefore, this formula comprehensively summarizes the influence of element amplitude and phase [5].

\begin{figure}[H]%调节图片位置，h：浮动；t：顶部；b:底部；p：当前位置
	\centering
	\includegraphics[width=0.82\linewidth]{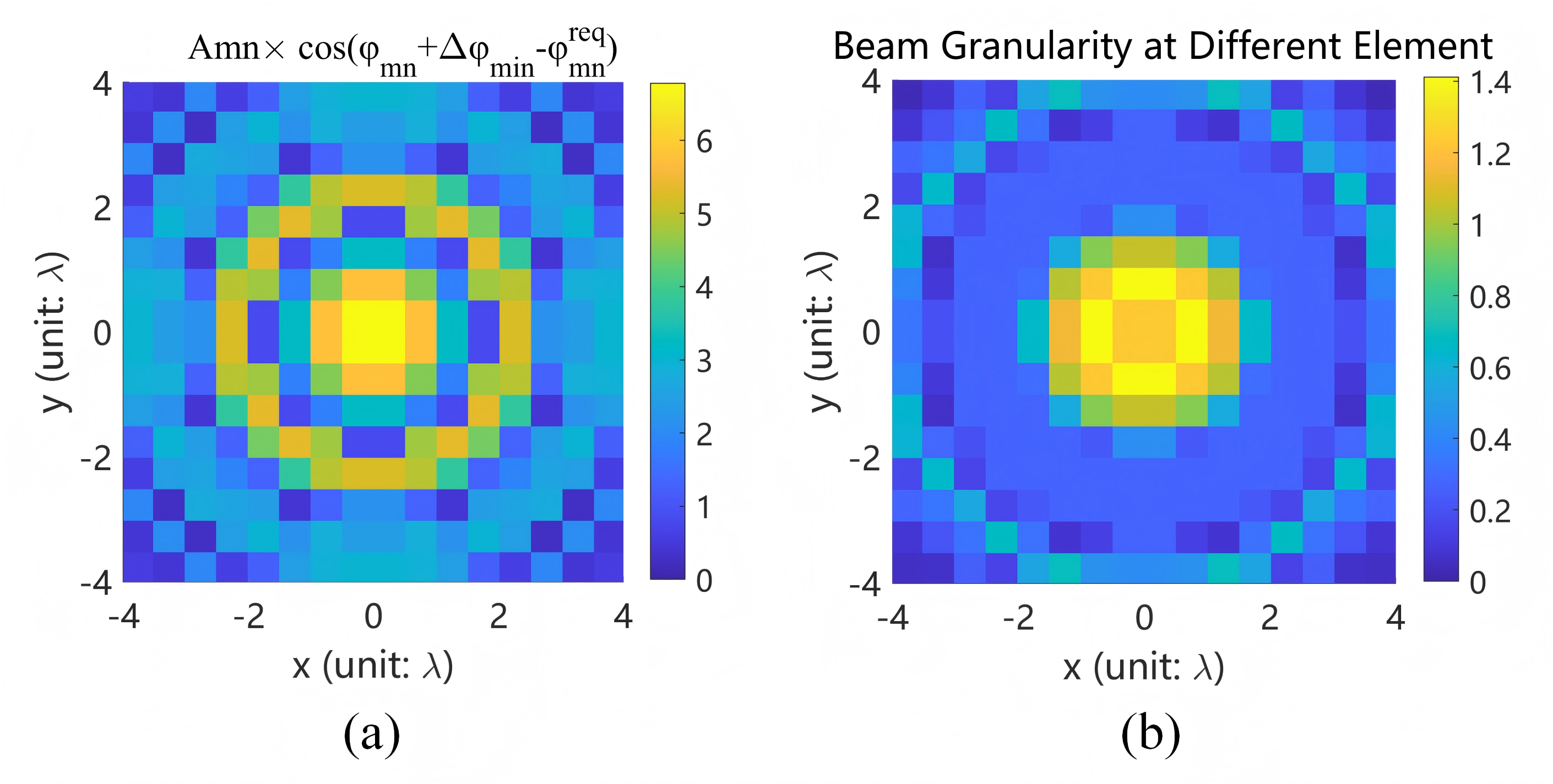}% 中括
	\caption{(a) Error distribution of the array based on the derived formula;; (b) Beam granularitys generated by units at different positions in the array.}
\end{figure}

After simulation verification, it can be found in Fig. 5 that the error distribution based on the derived formula is basically consistent with the beam granularity distribution, and the position with the minimum error is the position with the minimum beam granularity.

 \section{Conclusions}
This article systematically studies the impact of spatial quantization and phase quantization on reconfigurable reflectarray radar beam granularity performance.  Through theoretical analysis and simulation verification, it is found that spatial quantization and phase quantization can both reduce the beam granularity of the array antenna. At the same time, we give the formula for finding the beam granularity distribution of reconfigurable reflectarray radar. This research holds significant reference value for applications in communication systems, radar detection and beam tracking.

\end{document}